\documentclass[11pt,a4paper]{article}
\usepackage[T1]{fontenc}
\usepackage{amssymb}
\usepackage{mathptmx}
\usepackage[pdftex]{graphicx}
\usepackage[english]{babel}
\usepackage[pdftex,linkcolor=black,pdfborder={0 0 0}]{hyperref}
\usepackage{calc} 
\usepackage{enumitem} 
\usepackage{setspace}
\usepackage[a4paper, lmargin=0.12\paperwidth, rmargin=0.12\paperwidth, tmargin=0.1111\paperheight, bmargin=0.1111\paperheight]{geometry} 
\usepackage{wallpaper}
\usepackage{mdframed}
\usepackage{tikz}
\usepackage{comment}
\usepackage[affil-it]{authblk}

\usepackage[all]{nowidow} 
\usepackage[protrusion=true,expansion=true]{microtype} 
\usepackage{csquotes}

\usepackage{xr} 
\makeatletter
\newcommand*{\addFileDependency}[1]{
  \typeout{(#1)}
  \@addtofilelist{#1}
  \IfFileExists{#1}{}{\typeout{No file #1.}}
}
\makeatother

\usepackage[backend=biber,style=chem-acs, autocite=superscript]{biblatex} 
\newcommand{\mat}[1]{$\mathrm{#1}$} 
\newcommand{\um}{$\mathrm{\mu m}\ $}
\newcommand{\ume}{$\mathrm{\mu m}$} 

\author[1,*]{Nicola Melchioni}
\author[1]{Luca Bellucci}
\author[2]{Alessandro Tredicucci}
\author[1,*]{Federica Bianco}

\affil[1]{NEST Laboratory, Istituto Nanoscienze CNR and Scuola Normale Superiore, Piazza San Silvestro 12, I-56127, Pisa}
\affil[2]{Istituto Nanoscienze CNR, Piazza San Silvestro 12, I-56127, Pisa and Dipartimento di Fisica "E. Fermi", Università di Pisa, Largo Bruno Pontecorvo 3, I-56127, Pisa}
\affil[*]{Corresponding authors: (N.M.) Tel. +39 050 509502, E-mail: nicola.melchioni@sns.it\\ (F.B.) Tel. +39 050 509378, E-mail: federica.bianco@nano.cnr.it}

\title{Operability timescale of defect-engineered graphene}

\hypersetup{ 	
pdfsubject = {},
pdftitle = {Operability timescale of defect-engineered graphene},
pdfauthor = {}
}

\begin{document}

\maketitle
\doublespacing

\section*{Abstract}
Defects in the lattice are of primal importance to tune graphene chemical, thermal and electronic properties. Electron-beam irradiation is an easy method to induce defects in graphene following pre-designed patterns, but no systematic study of the time evolution of the resulting defects is available. In this paper, the change over time of defected sites created in graphene with low-energy ($\leq 20$ keV) electron irradiation is studied both experimentally via micro-Raman spectroscopy for a period of $6\times 10^3$ hours and through molecular dynamics simulations. During the first 10 h, the structural defects are stable at the highest density value. Subsequently, the crystal partially reconstructs, eventually reaching a stable, less defected condition after more than one month. The simulations allow the rationalization of the processes at the atomic level and confirm that the irradiation induces composite clusters of defects of different nature rather than well-defined nanoholes as in the case of high-energy electrons. The presented results identify the timescale of the defects stability, thus establishing the operability timespan of engineerable defect-rich graphene devices with applications in nanoelectronics. Moreover, long-lasting chemical reactivity of the defective graphene is pointed out. This property can be exploited to functionalize graphene for sensing and energy storage applications.\\

\textbf{Keywords}: graphene; electron-irradiation; structural defects; operability timescale; micro-Raman spectroscopy; time stability; molecular dynamics simulations;

\section{Introduction}
Graphene in its pristine form has been extensively studied for its outstanding mechanical, physical, and chemical properties and is employed in electronics, optoelectronic devices \cite{haibo_optoelectronics}, energy storage \cite{en_st1}, chemical sensing \cite{patil_chemsens}, and water filtering \cite{boretti_desal1}. Although the presence of defects in the lattice is usually undesired, as it lowers the electronic mobility and broadens the phonon dispersion relation \cite{Banhart2011}, an antipodal approach makes use of defects intentionally created in the graphene sheet to tailor its chemical, thermal, electronic, and mechanical properties and realize devices with novel functionalities \cite{Liu_def2}. As an example, structural defects strongly enhance the surface chemical reactivity of graphene both in the presence (vacancy-type defects) or the absence (topological-type defects like Stone-Wales and reconstructed vacancies) of dangling bonds \cite{Yang2018,Boukhvalov2008, Hernandez2013,Ye2017}. This allows the development of novel chemically functionalized devices for sensing, energy storage and energy harvesting applications \cite{lee_sens1, sunnardianto2021_hydrostorage1, battacharya_hydrostorage2, liang_energystorage2, akilan_energystorage}. Additionally, mastering graphene defects would have strong impact also in solid-state quantum technologies. In particular, lattice defects can be exploited for engineering the graphene charge and thermal properties and achieving full control of the energy and charge transport. Indeed, defects may transform graphene in an electrical insulator due to the reduced electrical conductivity \cite{Liu2011}, open a local bandgap (up to 0.3 eV) or bring the electrons into strong localization regime \cite{Yang2018}. Moreover, the weaker carbon bonds around defects strongly reduce the thermal phonon conductivity, up to one order of magnitude \cite{Malekpour2016}, and enhance about 3 times thermoelectricity at room temperature \cite{Anno2017}. All these altered properties can lay the foundation for the realization of new device concepts, like memories, logic gates, radiation detectors \cite{Han2013} and thermoelectric elements.\\
In general, all these devices require both fine tailoring and time stability of the modified properties for real-world applications. The former is guaranteed by the precise control of the density and spatial distribution of the defects. For example, an efficient method allowing to expose chosen patterns with nanometer precision is based on energetic electron irradiation \cite{Yang2018,liu_ebeam3,li_ebeam4, sun_lowenergy1, murakami_damage1, gu_metaldamage, basta2021_substrate, xin_holes1, twelde_modification1}. On the contrary, the latter depends on the time stability of the induced defects and their interaction with the environment. Generally, the high formation energy of defects ($> 1\ \mathrm{eV}$, depending on the nature of the defect \cite{Banhart2011}) should ensure the stability over time also at room temperature. Partial lattice healing can be achieved by annealing the sample in vacuum or in a controlled atmosphere but a layer of amorphous carbon forms on the surface of graphene due to organic residues \cite{shlimak_heal2, liu_graphitic}. However, energy barriers for the defect migration are sufficiently low and lattice reconstruction can occur even at room temperature \cite{Liu_def2}. For example, vacancies can be healed by successive interaction of graphene with CO and NO molecules that fill the hole and remove oxygen from the structure, respectively \cite{wang_heal1}. In addition, C adatoms can cluster and create mobile structures that induce self-healing of hillock-like defects \cite{Tsetseris_heal}. Indeed, modification during time of defects induced by particle irradiation was observed in suspended graphene where carbon atoms were present as hydrocarbon contaminants \cite{zan_stab2}. A partial self-healing of the defects was measured in systems where defects were induced in CVD-grown graphene by e-beam chemistry with water vapour in a scanning electron microscope (SEM) \cite{ISLAM_stab1}. Despite these many observations, it is not yet clear what is the complex process through which such induced defects heal and their actual time stability over a long period (e.g. months-scale) when controlled healing processes are not applied and the defective graphene ages at room-temperature and in ambient air conditions, as it may happen in defect-rich graphene-devices.\\
In this paper, we study the long-period time stability of electron-irradiated graphene when supported by the standard silicon dioxide/silicon substrate of device physics. In particular, we show that, when modifying the graphene lattice by low-energy (i.e. 20 keV) electron beam irradiation (EBI), the induced defects evolve over time at room temperature through a partial self-healing of the crystal, until they finally reach a stable and less defected state in about 700 h. Moreover, a change over time in the nature of the defects is observed. The population shifts from a majority of vacancy-like defects towards a majority of $sp^3$-like atoms. The time evolution was experimentally investigated by micro-Raman spectroscopy for $6\times 10^3$ h.\\
To understand the evolution of defected graphene over time at atomistic level, simulations at atomistic level were performed by exploiting molecular dynamics (MD) \cite{Senftle_2016, Lin_2015}. The recent parametrization of reactive empirical bond-order force field (ReaxFF) \cite{van2001reaxff,kowalik2019,kowalik2021atomistic, Lin_2015, marsden_2022, ozden_2022} and a revised primary knock-on atom (PKA) approach were profitably employed to model the effects of electron irradiation at low energy \cite{xia2016EBI, jang2004EBI, asayama2012EBI, kida2015EBI}. It is worth noting that the simulations intend to statistically study the general temporal behavior of a wide range of possible defects that can be experimentally generated. To this scope, the adopted simulation methods represent the most suitable approach (stochastic, unbiased and ensuring high statistics) to simulate the EBI effects on the graphene lattice. Consequently, the simulations mimic the experiments, reproducing the main defect features, but they do not intend to achieve an exact correspondence among simulated and experimental defective sites. The simulations reproduced the same trends observed in the experimental data, demonstrating that, once local lattice damage includes vacancy defects with chains, cross-linking, and distorted C atoms,  the relaxation of such defected systems displays a self-healing of the lattice. Moreover, we also probed the reactivity of the defected carbon atoms originated after the irradiation by simulating the systems in the presence of atomic hydrogen (H). Interestingly, the simulations revealed an increased reactivity with respect to pristine graphene also on long timescales, which is important in the case of graphene functionalization processes.

\section{Results and discussion}
\begin{figure}[t]
    \centering
    \includegraphics[width=1\textwidth]{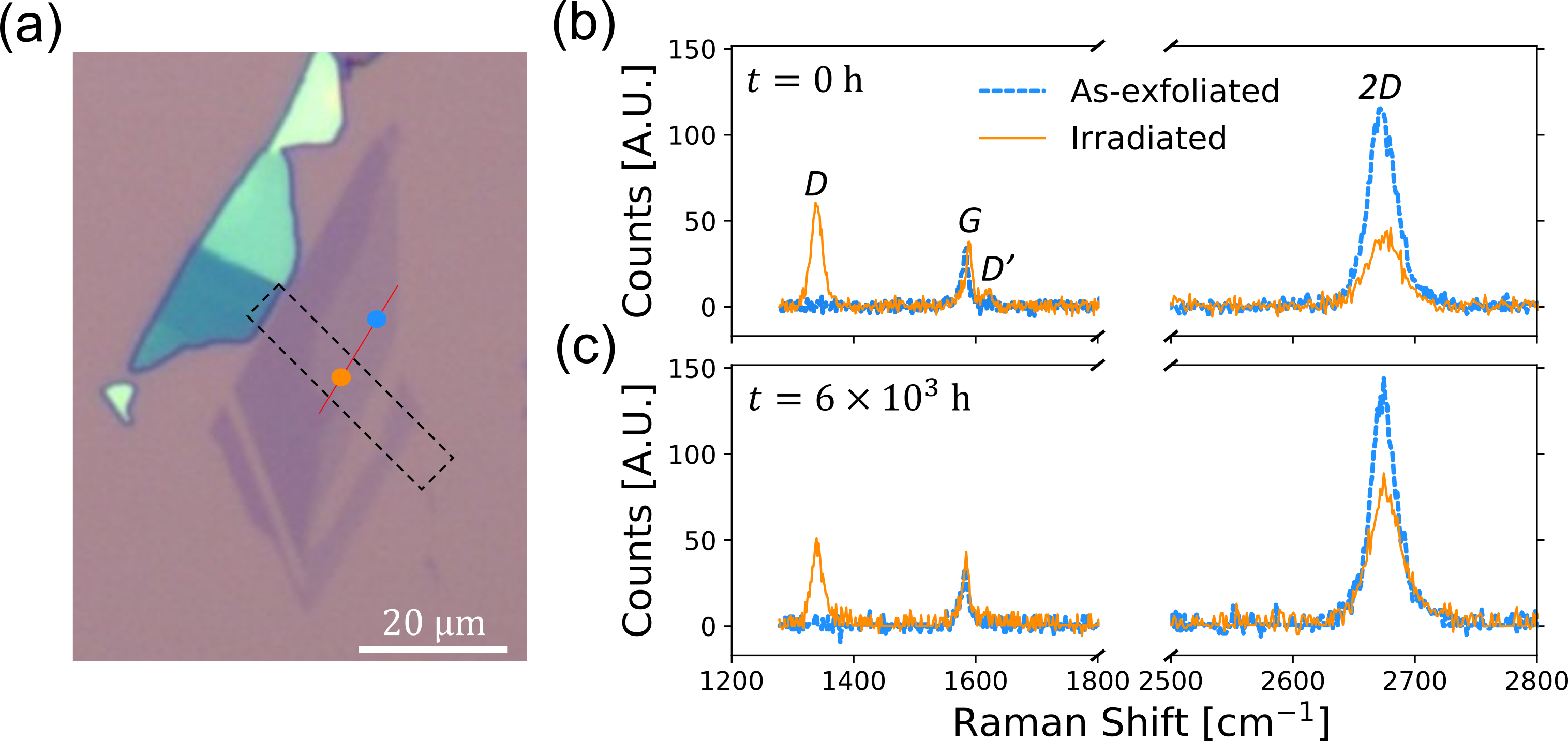}
    \caption{(a) Optical image of the graphene flake on \mat{SiO_2} substrate used for the presented experiment. The electron beam was scanned over the area enclosed in the black dashed rectangle which is 6 \um long and as wide as the flake. The micro-Raman laser was scanned along the red continuous line to collect spectra with 1 \um resolution. (b) Raman spectra collected on as-exfoliated graphene (dashed blue line) and on irradiated graphene (continuous orange line) right after the treatment at a time $t = 0\ \mathrm{h}$ (above) and after aging in air atmosphere at $t = 6\times 10^3\ \mathrm{h}$ (below).
    }
    \label{fig1}
\end{figure}

\begin{figure}[t]
    \centering
    \includegraphics[width=1\textwidth]{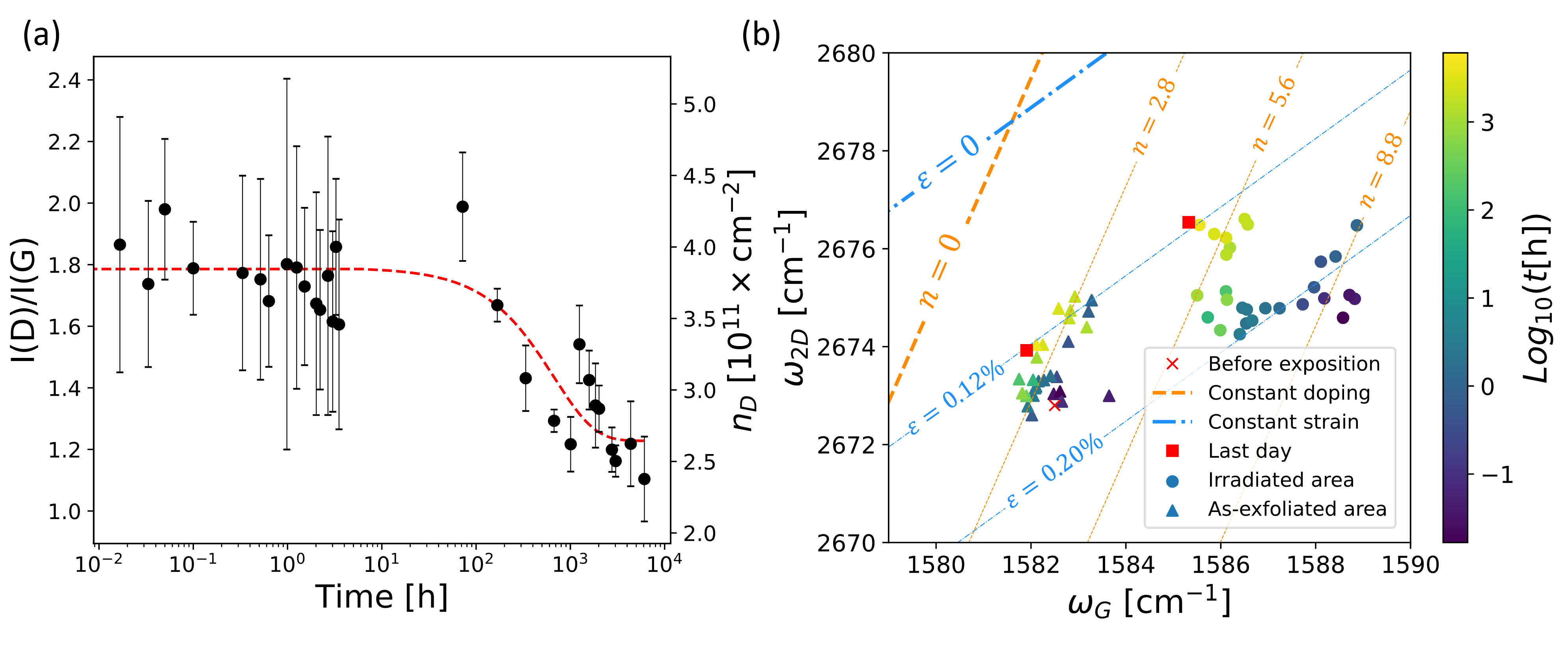}
    \caption{(a) Evolution over time of the ratio $I(D)/I(G)$ in the spectra collected on the irradiated area of graphene. The error bars are associated to statistical error over multiple consecutive measurements on the same line. The red dashed line is the result of a fit with an exponential decay. From the fit, a decay time $\tau_D = 670\ \mathrm{h}$ and a percentage of healed defect $\rho_{hD} \sim 31 \%$ are extracted. (b) Correlation of 2D peak position $\omega_{2D}$ versus G peak position $\omega_G$ for as-exfoliated (triangles) and irradiated (circles) graphene. The color bar represents the base-10 logarithm of the elapsed time, measured in hours. Lines of equal strain $\varepsilon$ (blue) and equal charge doping $n$ (orange) are reported for different conditions (see Ref. \cite{lee2012}). Doping values are given in units of $10^{12}\ \mathrm{cm^{-2}}$.
    }
    \label{fig2}
\end{figure}

The sample, shown in Figure \ref{fig1}a, was characterized before the irradiation to confirm the high quality of the monolayer graphene crystal (See SI). The sample was then irradiated as described in methods to induce defects in the lattice. Figure \ref{fig1}b shows the comparison between Raman spectra collected on irradiated and as-exfoliated graphene right after the exposure to the e-beam ($t_0 = 0 \ \mathrm{h}$). The spectrum collected on the as-exfoliated part is very similar to the one collected on pristine graphene (Figure S4), demonstrating that the crystal lattice of the unexposed area is not affected by the irradiation process. Instead, the spectrum of irradiated graphene features the typical disordered graphene fingerprint, i.e. a quenched $2D$ peak and a $D$ peak higher than the $G$ peak \cite{krauss2009, Wu_2018}. Also, the $G$ peak is blue-shifted with respect to the same peak on pristine graphene and a $D'$ peak is visible. See Table S3 for the values obtained from the Lorentzian fits of the peaks. Such a difference in the spectra confirms the presence of defects locally induced by the irradiation. To quantify such defects, the ratio $I(D)/I(G)$ of the intensities of the $D$ and $G$ peaks is used. A ratio of $I(D)/I(G)|_{t_0} = 1.86$ is observed at $t_0$. This ratio can be linked to the density of defects in the lattice  by the formula \cite{Bruna2014}:
\begin{equation}
    n_D [\mathrm{cm^{-2}}] = 7.3\times10^9 E_L^4[\mathrm{eV^4}]\frac{I(D)}{I(G)}, 
\end{equation}
where $E_L$ is the Raman laser excitation energy. By using this formula, the observed $I(D)/I(G)$ ratio results in a starting density $n_D(t=t_0) \sim 4\times 10^{11}\ \mathrm{cm^{-2}}$. Figure \ref{fig1}c shows the comparison between the Raman spectra collected on the same irradiated and as-exfoliated graphene after a time $t_f=6\times 10^3\ \mathrm{h}$ from the exposition. The signal collected on as-exfoliated graphene is comparable to the spectrum at $t_0$, showing that pristine graphene did not deteriorate during time. Conversely, the spectrum of irradiated graphene shows a reduction of the $D$ peak and a partial restoration of the intensity of the $2D$ peak. The $G$ peak is shifted closer to the as-exfoliated condition and the $D'$ peak is reduced (see Table S3). The intensity ratio is $I(D)/I(G)|_{t_f} = 1.1$, corresponding to a density of defects at the end of the experiment $n_D(t=t_f) \sim 2.3\times 10^{11}\ \mathrm{cm^{-2}}$ \cite{Bruna2014}. \\
To study the time evolution at room temperature of the induced defects, different Raman measurements were performed over time on the device. Figure \ref{fig2}a shows the $I(D)/I(G)|_t$ ratios extracted from spectra taken at different time $t$ on the irradiated part of the flake. The corresponding values of $n_D$ are reported on the right $y$-axis. Two different phases of the evolution can be distinguished: a first phase (from $10^{-2}$ h to $10$ h in Fig. \ref{fig2}a), where the density of defects is almost constant over a timescale of 10 hours, and a second phase (from $10$ h to $10^4$ h), in which a reduction of the density is observed in a timescale of $10^2-10^3$ hours, suggesting the occurrence of a room temperature healing process. To better quantify the timescale of the $I(D)/I(G)$ reduction, the data were fitted with an exponential decay (see SI and Figure \ref{fig2}a). A time constant of $\tau \sim 670$ h was extracted, while the percentage of defects that heal during the evolution is quantified to be as high as $\rho_{hD} = 1-(n_D(t_f)/n_D(t_0)) =  31\%$.\\ 
In the case under analysis, the defect density is low enough to still consider graphene as nanocrystalline \cite{eckmann_fwhmd, lucchese2010}. In such regime, the positions $\omega_G$ and $\omega_{2D}$ of the $G$ and $2D$ peaks can be correlated to extract information on doping and strain of the examined crystal structure (Figure \ref{fig2}b) \cite{lee2012}. At time $t_0$, the irradiated part shows a higher doping with respect to the as-exfoliated area. This is expected due to charge impurities accumulated in the substrate during the exposition \cite{park2009}. Also, the defected area shows a higher tensile strain than the as-exfoliated area, due to the presence of defects \cite{Leyssale_MDstrain1, blanc_strain}.
Instead, the data collected on the as-exfoliated area are comparable to the values obtained on the crystal before irradiation (red cross in Figure \ref{fig2}b). Importantly, the evolution over time of the two areas is different. In the irradiated area, two temporal phases can be isolated as in Figure \ref{fig2}a. During the first 10 hours, the doping decreases from a value $n \sim 8.8 \times 10^{12}\ \mathrm{cm^{-2}}$ to $n \sim 5.6 \times 10^{12}\ \mathrm{cm^{-2}}$. This change is attributed to the decay of the charge impurities accumulated in the substrate during the exposition. Indeed, charge impurities in \mat{SiO_2} have been observed to last up to few hours ($\lesssim 10$ h) after the irradiation \cite{burson2013}. After the decrease in doping, the crystal lattice relaxes its tensile strain from a value $\varepsilon = 0.20\%$ to a value $\varepsilon = 0.12\%$ in a timescale of $10^2-10^3$ hours. During the relaxation, the doping change is negligible. Instead, the as-exfoliated area follows a different path. First, the evolution cannot be divided into two distinguishable phases as in the case of the irradiated area. Second, the strain in the lattice decreases over time while the doping remains constant at $n \sim 2.8 \times 10^{12} \ \mathrm{cm^{-2}}$. The two examined parts eventually reach a condition where both share the same value of strain $\varepsilon = 0.12\%$. \\
\begin{figure}[t]
    \centering
    \includegraphics[width=1\textwidth]{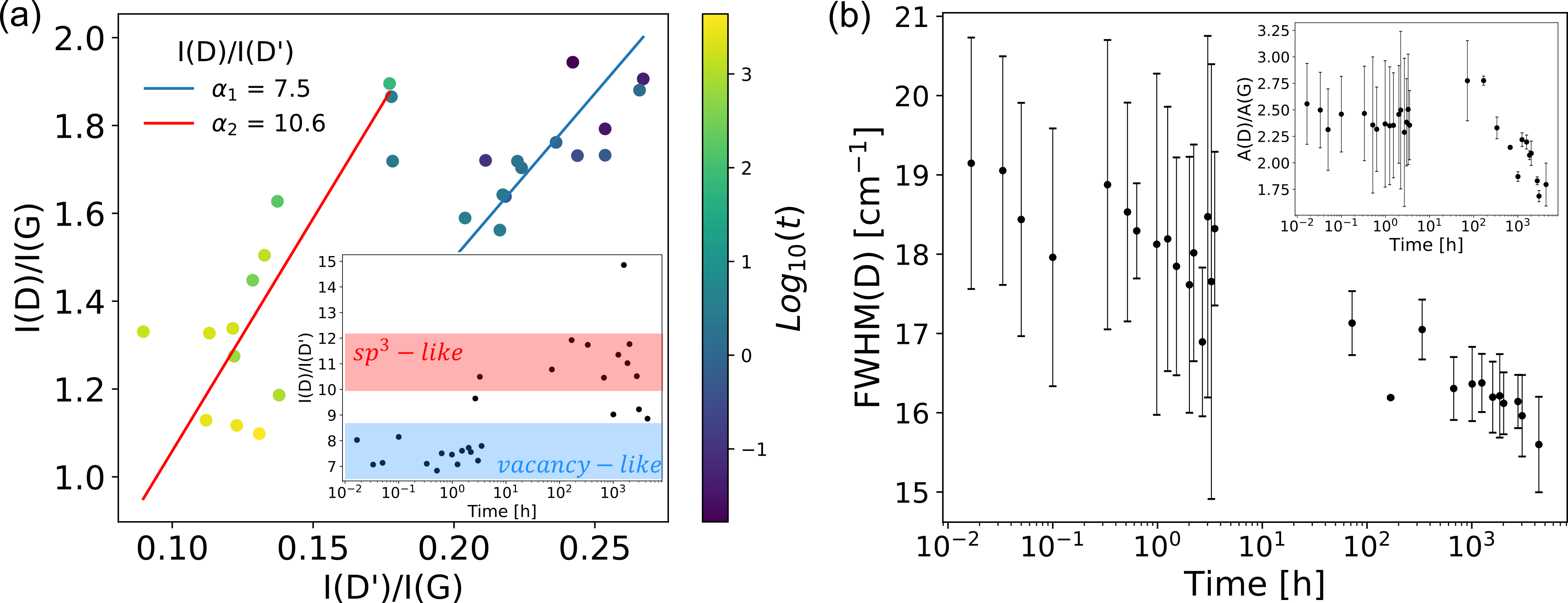}
    \caption{(a) Plot correlating the intensities of $D$ and $D'$ peaks collected on the irradiated area of the flake, normalized to the intensity of the $G$ peak. The color bar represents the base-10 logarithm of the elapsed time, measured in hours. Linear fits of the two temporal phases characterizing the evolution are shown (solid lines). In the inset, the ratio $I(D)/I(D')$ is plotted as a function of time for a better look at the time evolution. (b) Plot of the time evolution of the full width at half maximum $FWHM(D)$ of the $D$ peak extracted from the spectra measured on the irradiated part of the flake. In the inset: plot of the time evolution of the ratio $A(D)/A(G)$ of the areas of the $D$ and $G$ peak, respectively. The error bars are associated to statistical error over multiple consecutive measurements on the same line.} 
    \label{fig3}
\end{figure}
To understand more in depth the mechanism driving the observed time evolution, we studied the nature of the defects by examining the intensities of $D$ and $D'$ peaks. A plot of the intensities of the two peaks normalized on the relative $G$ peak is shown in Figure \ref{fig3}a. A linear correlation between the two parameters is visible with a coefficient $\alpha = I(D)/I(D')$. In general, this ratio gives information on the statistical composition of the defect population \cite{eckmann_dprime}. Interestingly, the two phases identified in the time evolution are characterized by two distinct $\alpha$-coefficients extracted by linear fits, also shown in Fig \ref{fig3}a. The first phase of the evolution is characterized by a coefficient $\alpha_1 \sim 7$. In the second phase, the extracted coefficient is $\alpha_2 \sim 11$. Assuming the model proposed in Ref. \cite{eckmann_dprime} for the $\alpha$-coefficient, the value of $\alpha_1$ indicates that, right after the exposition, the defect population is mainly composed of vacancy-like sites, such as borders and holes in the lattice. In the case of electron-nucleus collisions, C atoms are displaced out of the graphene lattice eventually creating atomic vacancies when incident electrons have energy larger than 86 keV, i.e. when the transferred energy exceeds the knock-on threshold energy \cite{krasheninnikov_holes2}. Nevertheless, lattice disruption are expected to occur also at 20-30 keV due to cumulative dose effect in continuous irradiation \cite{xin_holes1, krasheninnikov_holes2} or due to electrostatic fields caused by charged puddles generated by the irradiation in case of substrates with dielectric films, such as SiO$_2$/Si \cite{Ilyin_2014, Ilyin_2013, Ilyin_2021}. 

After the creation of the defects, vacancies-rich defects remain in the first $10$-hour phase (blue-shaded area in the inset of Figure \ref{fig3}a). Then, the $\alpha$ coefficient increases up to the $\alpha_2$ value. This indicates a change of the population towards a majority of $sp^3$-like defects (red-shaded area in the inset of Figure \ref{fig3}a). In addition, air contaminants can bond to the dangling bonds at the boundaries of the holes, thus changing the hybridization of the involved carbon atoms. This process, combined with the behaviour of $I(D)$ shown in Figure \ref{fig2}a, might suggest that the defect density does not decrease and the healing process is in fact absent. Indeed, if many contaminants bonded to the carbon dangling bonds, the measured intensity of the $D$ peak would be quenched by the presence of such molecules \cite{eckmann_fwhmd}. Hence, the density of defects would not actually decrease. Moreover, the full width at half maximum $FWHM(D)$ of the $D$ peak would increase, because of the larger energy spectrum of the bonds between carbon and different external molecules \cite{eckmann_fwhmd}. Albeit some external molecules may be bonded to the defective sites due to their enhanced chemical reactivity, in our sample $FWHM(D)$ narrows with time. As shown in Figure \ref{fig3}b, the width of the D peak decreases from $19 \ \mathrm{cm^{-1}}$ to a final value of $16 \ \mathrm{cm^{-1}}$. Such a decrease implies that the area below the measured $D$ peak diminishes over time, as shown in the inset of Figure \ref{fig3}b, and thus, that a decrease of the density of defects in the crystal takes place. Consequently, the $FWHM(D)$ evolution, combined with the partial 2D peak restoration (see SI Figure S5), confirms the presence of a healing process.\\
\begin{figure}[t]
    \centering
    \includegraphics[width=1\textwidth]{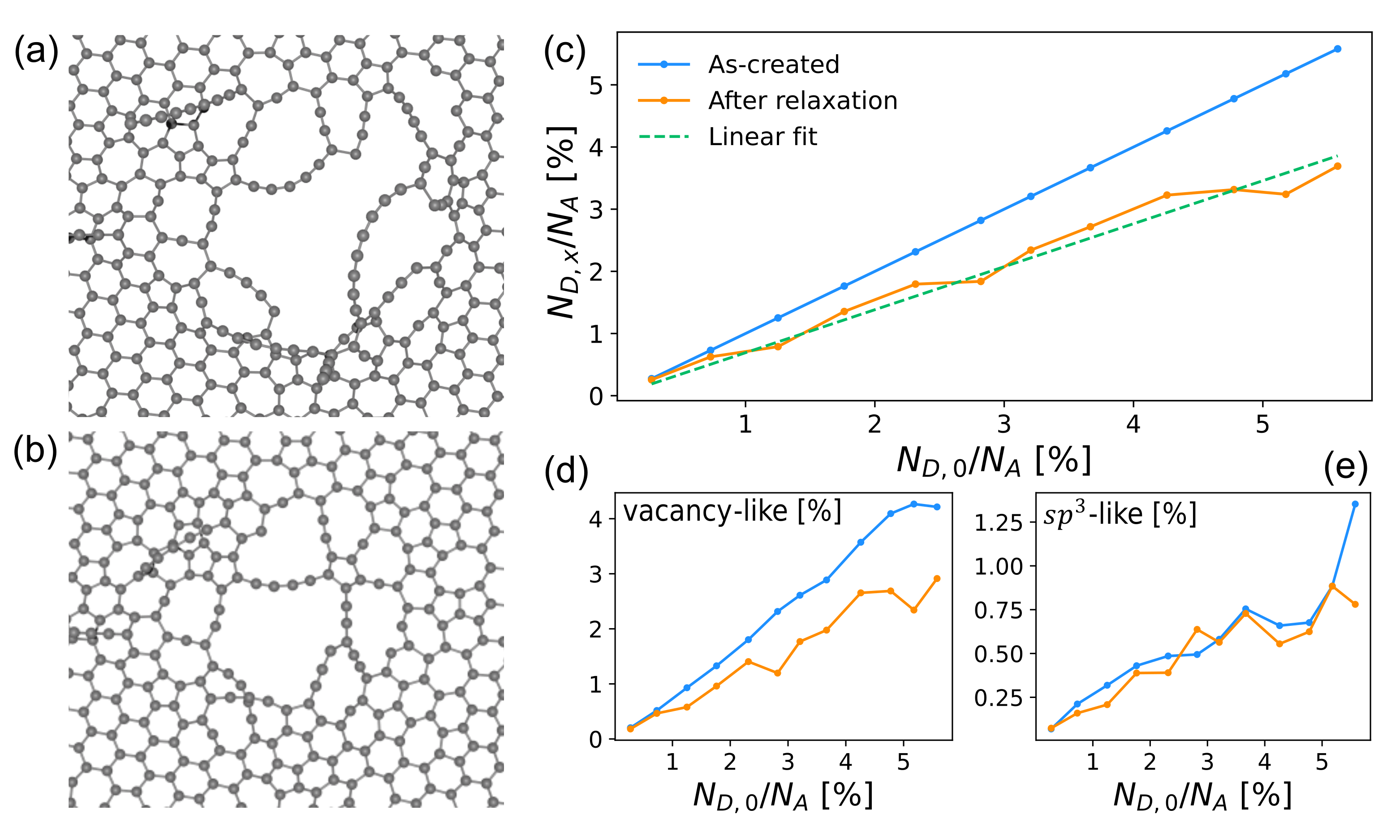}
    \caption{(a) Example of a simulated defected site. (b) The same site as in (a) after the relaxation process in vacuum. (c) Percentage of defected atoms in the system before (blue line, $x = 0$) and after (orange line, $x = f$) the relaxation process in vacuum as a function of the percentage of defected atoms before the relaxation. The green dashed line is a linear fit. The extracted coefficient is $\rho \sim 0.7$. (d, e) Percentage of vacancy-like (d) and $sp^3$-like (e) defects before (blue line) and after (orange line) the relaxation process in vacuum.} 
    \label{fig4}
\end{figure}
To explain the observed behaviour on the atomic scale, MD simulations were performed to mimic low-energy EBI (see Methods section). Figure \ref{fig4}a and b show an example of such created defected sites right after the end of the irradiation process and after relaxation, respectively. As-created defects (Figure \ref{fig4}a) appear like irregular defects composed of vacancies, carbon polygons and carbon chains, rather than a well defined nanopore. During the irradiation, the planar graphene structure is partially converted into a corrugated lattice with $sp$, $sp^3$, and unsaturated C atoms (see Figure S6 in SI for further examples). During the relaxation, unsaturated atoms tend to be saturated by other C atoms. Thus, the defected site reduces in lateral size and parts of the crystal lattice are reconstructed (Figure \ref{fig4}b). Several equivalent systems were irradiated to create different defected sites with increasing number of amorphous C atomic structures. Figure \ref{fig4}c shows the percentage of defected atoms ($N_{D,x}$) over the total number of atoms in the systems ($N_A$) (see SI) before ($x=0$) and after the relaxation ($x=f$) in vacuum as a function of the percentage of defected atoms in the as-created defected systems ($N_{D,0}/N_A$). The blue line referring to the as-created conditions can be used as a reference. Interestingly, ($N_{D,f}/N_A$) shows a linear behaviour (orange curve in Figure \ref{fig4}c) with an extracted linear coefficient $\rho = 0.7$. This suggests that the healing process depends on the size of the defected area but, in average, the fraction of healed defects is at least about $30\%$ of the starting distorted atoms, in good agreement with the experimental observations. As seen in Figure \ref{fig3}a, vacancy-like defects can be distinguished from $sp^3$-like defects by a different Raman activity. However, a more detailed definition can be given from the atomistic point of view. In fact, vacancy-like defects and $sp^3$-like defects are composed by different kinds of C atoms, which can be identified in the simulations (see SI). The first group includes atoms at the border of large holes or linked to linear chains, like the one represented in Figure \ref{fig4}a. The second group includes $sp^3$-hybridized atoms and quantifies the deviation from the planarity of $sp^2$-hybridized C in the lattice.
The results on the simulated evolution of these two groups allow to better understand the experimentally observed healing process, i.e. the change from vacancy-rich to $sp^3$-like population. The number of vacancy-like defects decreases during relaxation as the hole partially refills with hexagonal rings (Figure \ref{fig4}d), while the $sp^3$-like population remains constant on average (Figure \ref{fig4}d). This behaviour can be explained as follows. On the one hand, the $sp^2$ atoms with dangling bonds that are at the edge of the hole and the linear carbon chains, i.e. the vacancy-like atoms, are prone to react towards saturation. Hence, by saturating the dangling bonds, the lattice is recovered and the structure is locally stabilized. Such a mechanism arises from the simultaneous availability of dangling bonds and movable linear C chains, which are intrinsic features of the created defects, and represent the driving force for the reconstruction of graphene. On the other hand, even if some distorted $sp^2$ atoms can relax towards a more planar disposition, other surrounding atoms may rearrange during the evolution as an effect of the reconstruction (see Figure \ref{fig5}a and b), thus explaining why the $sp^3$-like population does not change on average.\\ 
\begin{figure}[b!]
    \centering
    \includegraphics[width=1\textwidth]{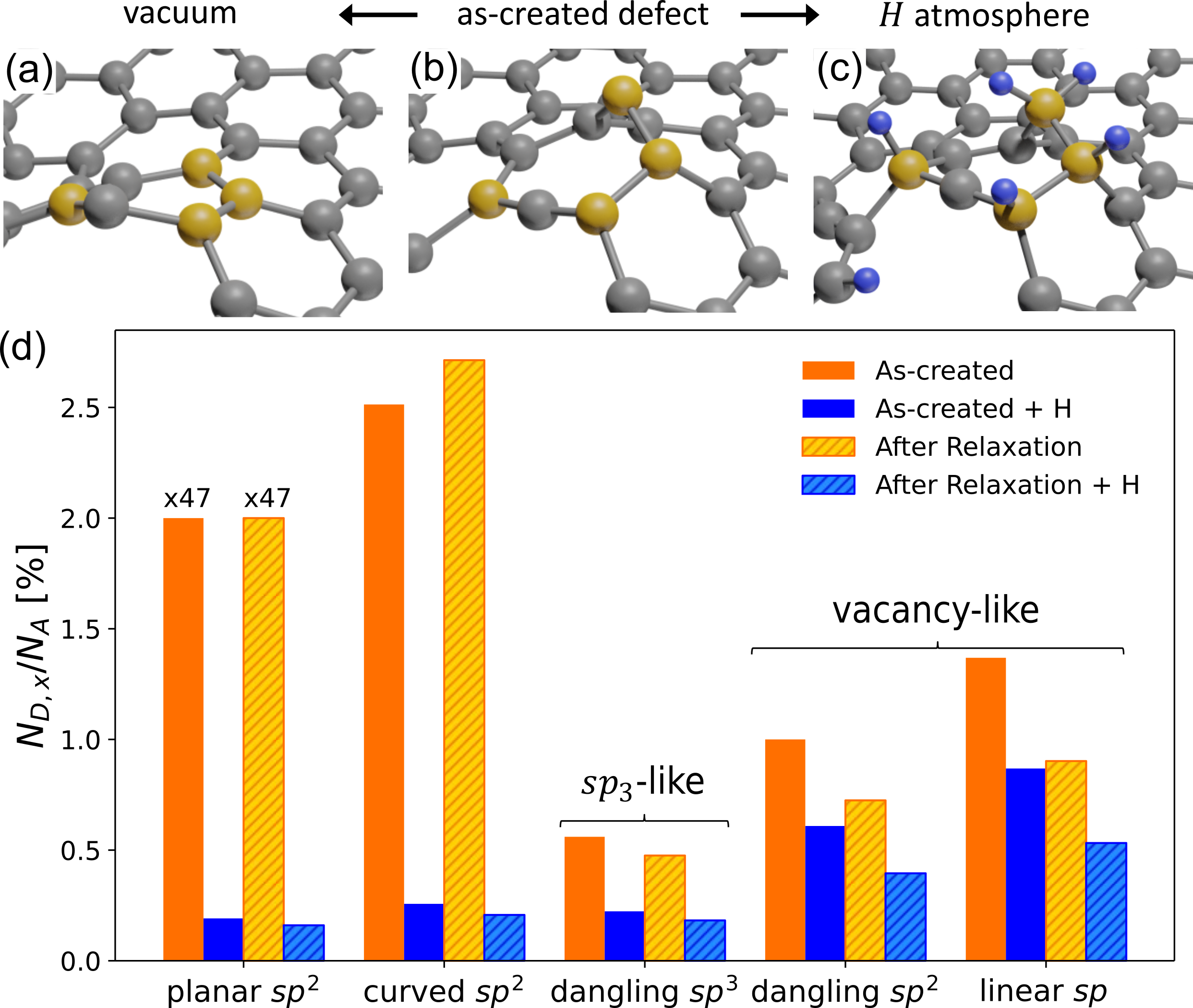}
    \caption{(a, b, c) Example of two possible evolution outcomes that a distorted $sp^2$ atom can undergo. The yellow spheres are the $C$ atoms in the lattice interested by such evolution, the blue spheres are $H$ atoms introduced as contaminants.
   (d) Histogram of the percentage of atoms categorized by hybridization and averaged over all the systems for the as-created systems (solid bars) and relaxed systems (textured bars). The orange bars show the average percentages of atoms before the hydrogenation, while the blue bars show the percentage of the atoms of the corresponding starting hybridization that bonded with hydrogen atoms.} 
    \label{fig5}
\end{figure}
The simulations in vacuum well reproduce the healing trends that are experimentally observed at microscopic level in ambient air. This can be explained by the fact that the lattice reconstruction involves defects with different nature, which evolve differently with time. In particular, when exposed to ambient air, defective sites may chemically react with airborne contaminants due to the enhanced reactivity\cite{Liu_def2, irbarne_def1}. Indeed, we observed a larger charge doping in the irradiated areas compared to the pristine graphene. However, the probability of chemical bonding between defects and airborne contaminants is modest at standard temperature and pressure \cite{Li_contaminants1}, as also demonstrated by the almost constant charge doping and the $FWHM(D)$ narrowing with time in our samples (see Figure \ref{fig2}a and Figure \ref{fig3}b, respectively). Consequently, the bonding of the defects with external molecules is not an immediate mechanism. This gives rise to a slow saturation process of the electron-induced dangling bonds, allowing the vacancy-like defects to be reconstructed during time.

To deeply investigate the role of the chemical reactive sites in the healing process, we carried out additional MD simulations and studied the graphene reactivity in the presence of defects. The systems were subjected to tempering cycles in the presence of atomic hydrogen (see Computational methods section). We employed atomic hydrogen in order to probe the reactivity of the defected sites with a very active species. As a first test, pristine graphene was simulated in H atmosphere. No atoms bonded to the surface (see video in SI), thus demonstrating that pristine graphene is chemically inert in the chosen condition (see Methods and SI). Instead, the defected sites interacted with H, yielding different outcomes. For example, if vacancy-like atoms are saturated by an H atom, the configuration likely stabilizes the $sp^2$ hybridization and that particular defect stops the healing mechanism observed in vacuum. If $sp^3$-like atoms are saturated by H atoms, the distorted $sp^2$ carbon atoms change their hybridization to $sp^3$ and the lattice out-of plane distortions convert into stable tertiary C atoms bound to H, as shown in Figure \ref{fig5}b and c. In general, H contamination of the as-created systems pointed out that the reactivity of planar $sp^2$ is not zero as in pristine (i.e. non-defected) graphene (see solid bars in Figure \ref{fig5}d). This implies that defects affect the reactivity of the system also at relatively longer distances. Curved $sp^2$ and $sp^3$-like atoms show an improved reactivity \cite{tozzini_curved1, ciammaruchi_curved2}, while dangling bonds show the highest reactivity (dangling $sp^2$ and linear $sp$ in Figure \ref{fig5}d). Besides, it is worth noting that, even when the systems were subjected to a strong chemical attack, the chemical saturation occurred gradually and some dangling bonds remain unsaturated by atomic H even after few aging cycles (see Figure \ref{fig5}d). Consequently, these results indicate that, even in the worst case scenario, i.e. the presence of high reactive species, the evolution of the system is a concurrency of the reconstruction mechanism of the lattice, which happens as if it were in vacuum, and the creation of distorted sites that can interact with possible contaminants and is responsible for the change in nature of the defects during time observed in the experiment (see Figure \ref{fig3}a).\\
For applications employing the defected sites in the functionalization of graphene, it is important that the improved reactivity is preserved over time. To test this, hydrogen attack was performed over the relaxed systems as well (textured bars in Figure \ref{fig5}d). After relaxation, the vacancy-like population decreases and consequently decreases the amount of saturated vacancy-like defects. Also in this case, the $sp^3$-like population remains constant during relaxation, guaranteeing an improved reactivity also after a long time.

\section{Conclusions}
The evolution over time of defects in a graphene flake was studied via both micro-Raman spectroscopy and MD simulations. Experimentally, the structural defects were patterned by low-energy EBI in an exfoliated graphene sheet. During the first $\sim 10$ h after the irradiation, the charge accumulated in the substrate due to the exposure relaxes and the local doping of graphene decreases, but the crystal modifications are otherwise unaffected and maintain the highest density value that is determined by the exposure parameters ($n_D(t_0) \sim 4\times 10^{11}\ \mathrm{cm^{-2}}$). 
On a much longer timescale of $\sim 700$ h, the actual crystal partially recovers and a stable defective state is maintained. This stable state is reached after a partial reconstruction of $\sim 30\%$ of the defects with a consequent local tensile strain relaxation, pointing out that graphene can rebuild its lattice when exposed to ambient air at room temperature without any specific healing treatment. The MD simulations results supported the observed reconstruction of the defected sites for at least $30 \%$ of the distorted atoms in vacuum. Moreover, the simulations showed that the induced defects are composite sites with vacancies and complex amorphous C structures and not well defined nanoholes. In fact, the experiment shows that vacancy-like defects are the main component of the defected site right after the exposure. Thereafter, such vacancy-like population decreases over time due to the reconstruction and to the possible saturation of the dangling bonds with airborne contaminants, while the $sp^3$-like population is on average constant over time due to rearrangements in the lattice. Interestingly, the simulations also showed that curved $sp^2$ atoms are generated by the irradiation and preserved during aging. Alike $sp^3$-like atoms, curved $sp^2$ atoms are more reactive than planar $sp^2$ carbon \cite{prato_nanotube} and curved graphene has been reported in the literature to be a suitable substrate for H chemisorption \cite{tozzini_curved1, ciammaruchi_curved2}. Therefore, the experiment combined with simulations highlighted that, when evolving during time, the modifications of the defected sites are a combination of self-reconstruction of the lattice and creation of reactive sites that promote the chemical interactions in case of defects-airborne contaminant collisions. Such a process occurs on a long timescale because defects must overcome an energy barrier to recombine in the crystal \cite{Banhart2011} and the probability of collision with airborne contaminants is modest \cite{Li_contaminants1}. \\
In summary, the presented work shows that the nature and density of defected sites created by low-energy EBI evolve during time at room temperature and in ambient air conditions. Our results are relevant when dealing with nanoelectronic devices, as they establish the operability timescale for applications. Moreover, when designing the device, the aging of the defective state, and thus the modification of the altered electronics properties, must be taken into account, even when the device is kept under vacuum. Indeed, the highest defective state lasts for a too limited time (less then one day). Then, the system undergoes a general properties modification for more than one month, until it reaches a less defective but stable state, which is more suitable for long-lasting devices applications.
A possible solution to have more stable as-created defects may be encapsulating the graphene sheet. However, additional investigation is necessary when considering this system, as the defect healing may still be non-negligible and its dynamics may be different. Finally, this work points out that defective graphene has chemical activity that lasts during time. The presented complete rationalization of the different self-healing paths and the study of the chemical reactivity of the irradiation-induced defects also provide the conceptual tools to develop advanced chemical functionalization strategies to build novel graphene-based devices for energy storage, sensing and catalysis.

\section*{Materials and methods}

\subsection*{Fabrication and characterization}\label{fabrication}
The employed substrate is a boron-doped \mat{Si} wafer with 300 nm of thermally-grown \mat{SiO_2} on top. The substrate was cleaned by oxygen plasma at 100 W for 5 minutes to remove organic residues on its surface. Graphene flakes were deposited on the substrate by micro-mechanical exfoliation of highly oriented pyrolytic graphite.\\
Defects were induced on part of a graphene flake by irradiating with electrons accelerated at 20 keV. The defected area was precisely targeted by a SEM driven by a pattern generator, and is 6 \um long and as wide as the graphene flake (see Figure \ref{fig1}a). The e-beam was set to output a current of about 0.15 nA and was then scanned with a step-size of 0.1 \ume. The e-beam delivered a dose of 40 \mat{mC/cm^2}, resulting in a dwell-time of 30 ms.The irradiation was carried out at the base operational vacuum of the SEM chamber, which is of the order of $10^{-7}$ mbar. The sample was also left to degas in vacuum inside the SEM chamber for one night. This precaution reduces at the minimum the presence of undesired contaminants during the processing. Additionally, the irradiation conditions (i.e. chamber vacuum and exposure parameters) guarantee a negligible deposition of a carbonaceous contaminant layer \cite{femioyetoro_2019}.\\
The induced defects were studied by micro-Raman spectroscopy in ambient air. The flake was scanned by a laser beam at 532 nm with 100x objective, giving a lateral resolution $<1$ \ume. The laser delivered a power of 118 \mat{\mu W}. A power so low was chosen to exclude any possible laser heating of the lattice, which could give contribution to the healing of the defects. Different measurements were taken in a temporal span of 250 days to study the time evolution of the defects. Each measurement consisted of line scans of the flake taken along the line depicted in Figure \ref{fig1}a. The signal collected on the defected part of the flake was isolated from the one collected on the pristine part via software. We checked with an automated script the presence of the D peak and averaged over the extracted points. Experimental errors were then represented by the standard deviation of the data from the average. The signal collected each time on the pristine part of the graphene flake was analyzed following the same method and used as a reference.

\subsection*{Computational methods}
\label{comp_methods}
All calculations were performed with the LAMMPS program \cite{LAMMPS}. Visual inspections of molecular dynamics (MD) trajectories, post analysis and figures were performed with VMD1.9.3 \cite{VMD}. The systems under study consisted of two layers of graphene, each placed on the two surfaces of an amorphous \mat{SiO_2} plate (\mat{aSiO_2}, see Figure S1) and treated as periodic. The chosen box is a triclinic cell with $a=b=7.6$ nm, $c=8$ nm; $\alpha=\beta=90^{\circ},\ \gamma=120^{\circ}$. The atomic interactions between the carbon atoms (C–C) in graphene and an hydrogen atom, where present, were treated with ReaxFF \cite{van2001reaxff,kowalik2019}. Interactions between graphene or H and \mat{SiO_2} were described by using Lennard-Jones potentials with the parameters derived from CHARMM force field \cite{cruz2006SiO2} (see SI). Further details about MD simulations can be found in SI.
EBI at 20 keV was simulated exploiting the primary knock-out atoms (PKAs) method in the framework of MD simulation.
A total of 38 equivalent systems were subjected to electron irradiation, thus generating 76 different defective graphene lattices (see SI, Figure S6). These irradiated systems were then subjected to a chemical attack by atomic hydrogen and a tempering/annealing procedure in vacuum by employing MD simulations. The chemical attack was performed by heating the system to 800 K and cooling it to 300 K in 10 ps 6 times in the presence of 1800 H atoms in the simulation box. The tempering/annealing procedure in vacuum was carried out by employing 10 cycles in which each system was heated to 3000 K and cooled to 300 K. Each cycle was 10 ps long. The relaxed systems were then subjected to the same H chemical attack procedure.

\section*{Authorship statement}
\textbf{Nicola Melchioni}: Data Curation, Formal Analysis, Investigation, Writing – Original Draft\\
\textbf{Luca Bellucci}: Data Curation (MD Simulations), Funding acquisition, Investigation, Methodology (MD Simulations), Writing – Review\&Editing\\
\textbf{Alessandro Tredicucci}: Funding acquisition, Writing – Review\&Editing\\
\textbf{Federica Bianco}: Conceptualization, Data Curation, Funding acquisition, Investigation, Methodology, Writing – Review\&Editing

\section*{Supporting Information}
Additional experimental details, computational and analysis methods (PDF); Movie of the simulated structures (MPG)

\section*{Data Availability}
All the data are available upon request to the corresponding authors.

\section*{Acknowledgments}
This research was partially founded by EU-H2020 FETPROACT LESGO (Agreement No. 952068) and by "MONstre-2D" Project under the PRIN 2017 call and "q-LIMA" Project under the PRIN 2020 call granted by the Italian Ministry of University and Research. The authors wish to acknowledge Dr. C. Coletti from Istituto Italiano di Tecnologia for the access to micro-Raman facility.

\newpage
\printbibliography

\end{document}